\documentstyle[12pt,epsf,epsfig]{article}
\def\beq{\begin{equation}}
\def\eeq{\end{equation}}
\def\bea{\begin{eqnarray}}
\def\eea{\end{eqnarray}}
\def\ba{\begin{array}}
\def\ea{\end{array}}
\def\mwph{m^{(ph)\, 2}_W}
\def\mwphS{m^{SM\, 2}_{W}}
\def\mzph{m^{(ph)\, 2}_Z}
\def\gph{g^{(ph)}}
\def\gphs{g^{(ph)\, 2}}
\def\gzph{g_Z^{(ph)}}
\def\calo{{\cal O}}

\newcommand{\NPB}[3]{\emph{ Nucl.~Phys.} \textbf{B#1} (19#2) #3}   
\newcommand{\PLB}[3]{\emph{ Phys.~Lett.} \textbf{B#1} (19#2) #3}   
\newcommand{\PRD}[3]{\emph{ Phys.~Rev.} \textbf{D#1} (19#2) #3}   
\newcommand{\PRL}[3]{\emph{ Phys.~Rev.~Lett.} \textbf{#1} (19#2) #3}

\def\gappeq{\mathrel{\rlap {\raise.5ex\hbox{$>$}}
{\lower.5ex\hbox{$\sim$}}}}
\def\permil{$\%\raise.20ex\hbox{$_0$}}
\def\lappeq{\mathrel{\rlap{\raise.5ex\hbox{$<$}}
{\lower.5ex\hbox{$\sim$}}}}
\begin{document}
\topmargin -1.0cm
\oddsidemargin -0.8cm
\evensidemargin -0.8cm
\pagestyle{empty}
\begin{flushright}
CERN-TH/99-47
\end{flushright}
\vspace*{5mm}
\begin{center}
{\Large\bf Effects of SM Kaluza-Klein excitations
}\\
\vspace{0.5cm}
{\Large\bf on electroweak observables}\\
\vspace{2cm}
{\large\bf Manuel  Masip$^{a,b}$ and 
Alex Pomarol$^{a,}$\footnote{On leave of absence from
IFAE, Universitat Aut{\`o}noma de Barcelona, 
E-08193 Bellaterra, Barcelona.}}\\
\vspace{.6cm}
{\it {$^{a}$Theory Division, CERN}\\
{CH-1211 Geneva 23, Switzerland}\\
\vspace{.4cm}
{$^{b}$Dpt. F\'\i sica Te\'orica y del Cosmos}\\
{Universidad de Granada}\\
{E-18071 Granada, Spain}\\}
\end{center}
\vspace{2cm}
\begin{abstract}
The presence of an extra dimension of size $R\approx$ TeV$^{-1}$
introduces a tower of 
Kaluza-Klein gauge boson excitations that affects the 
standard model (SM) relations between electroweak observables. 
The mixing of the $W$ and $Z$  bosons with their excitations
changes  their masses and  couplings to fermions.
This effect depends on the Higgs field,
which may live in the bulk of the extra dimension, on its
boundary, or may be a combination of both types of fields.
We use high-precision electroweak data to constrain $1/R$. We
find limits from 1 to 3 TeV from different observables, with
a model independent lower bound of 2.5 TeV.
\end{abstract}

\vfill
\begin{flushleft}
CERN-TH/99-47\\
February 1999
\end{flushleft}
\eject
\pagestyle{empty}
\setcounter{page}{1}
\setcounter{footnote}{0}
\pagestyle{plain}


\section{Introduction}
 
It is widely believed that the standard model (SM) is 
the low-energy limit of a more fundamental theory including
gravity. It is also believed that 
one of the requirements for  this fundamental theory 
is the existence of more than three spatial dimensions, which
would be compact and with a radius $R$ of Planckean size.
 Recently, however, it has been suggested that the 
extra dimensions  can appear at much lower energies. 
A first possibility 
was  given in Refs.~\cite{a,ab} in the context of string theory.
It was shown that large extra dimensions 
do not necessarily spoil the
gauge coupling unification  of the 4D theory.
A more radical possibility, proposed in Ref.~\cite{add},
is to decrease the scale of unification of gravity with
the gauge interactions down to the TeV. 
This can be realized  by means of two sub-millimeter extra dimensions 
in which only gravity propagates.
Although the gauge interactions would not feel these sub-millimeter
dimensions, a fundamental scale (string scale) in the TeV
region suggests the possibility of compact dimensions 
of this size where the SM fields do propagate. 

Large ($\approx$ TeV$^{-1}$) extra dimensions 
find also an interesting motivation 
as a framework to break supersymmetry \cite{ss}.
This has been studied in  detail in Refs.~\cite{pq,adpq,dpq}, 
where it was predicted a compactification scale around $3-20$ TeV.
Also recently, it was discussed \cite{ddg,zurab} how
an extra dimension could lead
to the unification of the gauge forces at the TeV-scale.

In this letter we study the effects of extra dimensions
on electroweak observables.  If the 
SM gauge bosons can propagate in a compact dimension, their (quantized) 
momentum along this dimension can be associated to the 
mass $n/R$ ($n=1,..,\infty$) of a 
 tower of Kaluza-Klein (KK) excitations.
As a  consequence the relations between electroweak observables
will be modified with respect to those of the 4D SM.
There are two kinds of effects. The first  one
is due to the presence of mixing between
the zero and the $n$-modes of the $W$ and $Z$ bosons.
This leads to a modification of the $W$ and $Z$ masses
and their couplings to the fermions.
The second effect arises from the exchange of 
KK-excitations of the $W$, $Z$ and $\gamma$ vector bosons, which
induces extra contributions to four-fermion interactions.
We calculate these effects and show how to put bounds
on the size of an extra dimension from high-precision 
electroweak data. 
We find limits on $1/R$ from 1 to 3 TeV from different observables, 
with a model independent lower bound of $\approx 2.5$ TeV.

\section{Framework}
 
The model that we want to study is based on an extension of the 
SM to 5D \cite{pq}.
The  fifth dimension $x^5$ is compactified on the segment 
$S^1/Z_2$, a circle of radius $R$ 
with the identification $x^5\rightarrow -x^5$.
This segment, of length $\pi R$, has two 4D boundaries at
$x^5=0$ and $x^5=\pi R$ 
(the  two fixed points of the orbifold $S^1/Z_2$).
The SM gauge  fields live in the 5D bulk,
while  the SM fermions are localized on the 4D boundaries.
The Higgs fields can be either in the 5D bulk or on the 4D boundaries.
Models with  the Higgs in the bulk have been considered in 
Refs.~\cite{pq,zurab}, while models with  Higgs on
the boundary  have been  considered in Refs.~\cite{ddg,dpq}.
The most general case consists of a SM Higgs which is a 
combination of both types of fields. We will then assume the presence of  
two Higgs doublets, $\phi_1$ and $\phi_2$, 
living respectively in the bulk and on the boundary.

To illustrate how to obtain the SM in such a framework (for more details
see Refs.~\cite{peskin,pq,dpq}), 
let us consider a U(1) gauge theory in 5D with two scalars, 
$\phi_1$ in the bulk and $\phi_2$ localized on the $x^5=0$ boundary,
together with a fermion $q$ living on the same boundary.
We assume that all these fields have U(1) charges equal to 1.
The  5D lagrangian is given by
\begin{equation}
{\cal L}_5=-\frac{1}{4g^2_5}
F_{MN}^2+|D_M\phi_1|^2+
\Big[
i\bar q\sigma^\mu D_\mu q+|D_\mu\phi_2|^2
\Big]\delta(x^5)\, ,
\end{equation}
where $D_M=\partial_M+iV_M$, $M=(\mu,5)$, 
and $g_5$ is the 5D gauge coupling.
The fields living in the bulk 
are defined to be even under the $Z_2$-parity, i.e., 
$\Phi(x^5)=\Phi(-x^5)$ for $\Phi=V_M,\phi_1$.
They can be Fourier-expanded as
\begin{equation}
\label{expansion}
\Phi(x^\mu,x^5)=\sum^{\infty}_{n=0}
\cos\frac{nx^5}{R}\Phi^{(n)}(x^\mu)\, .  
\end{equation}
Using Eq.~(\ref{expansion}) and 
integrating  over the fifth dimension, the resulting  4D theory
(in the unitary gauge \cite{dpq}) is given by
\begin{eqnarray}
{\cal L}_4&=&\sum^{\infty}_{n=0}\Bigg[
-\frac{1}{4}
F^{(n)\, 2}_{\mu\nu}+{1\over 2}
\left(\frac{n^2}{R^2}+2g^2|\phi_1|^2\right)
V^{(n)}_\mu V^{(n)\, \mu}
\Bigg]\nonumber\\
&+&
g^2|\phi_2|^2\Bigg(
V^{(0)}_\mu + \sqrt{2}\sum^{\infty}_{n=1}
V^{(n)}_\mu \Bigg)^2
+
i\bar q\sigma^\mu \left[\partial_\mu+igV_\mu^{(0)}+ig\sqrt{2}
\sum^{\infty}_{n=1}V_\mu^{(n)}
\right] q+\dots\, ,
\label{4dlagrangian}
\end{eqnarray}
where  $g$ is now the 4D gauge coupling, related to the 5D coupling
by $g=g_{5}/\sqrt{\pi R}$.
We are only writing the terms which are relevant to generate 
gauge boson masses via Higgs 
vacuum expectation values (VEVs) and the couplings of the 
gauge KK-excitations $V^{(n)}_\mu$ 
to the fermions on the boundary.
These are the only types of terms that will be needed in our 
analysis. Two comments are in order. 
Due to the presence of the boundary
field $\phi_2$ 
and its VEV, the zero and $n$-mode of the gauge 
boson will mix. 
The mixing terms are allowed due to the breaking of
$x^5$-translational invariance by the boundaries.
Second, the coupling of the KK-excitations to the  fermion is enhanced
by a factor of $\sqrt{2}$ due to the different normalization of the zero
and the $n$-modes in the KK-tower.

The generalization of the above lagrangian to the SM is straightforward.
Following the standard notation, we will parametrize the VEVs of the Higgs by 
$\langle\phi_1\rangle= v\cos\beta\equiv vc_\beta$ and
$\langle\phi_2\rangle= v\sin\beta\equiv vs_\beta$
\footnote{We do not specify the couplings of $\phi_1$ or $\phi_2$ 
to the fermions since it is not needed here. 
However, the fact that the coupling of $\phi_1$ to the boundary is suppressed
by a factor $\sqrt{\pi R}$ suggests that $\phi_1$ ($\phi_2$) 
is the responsible for giving mass to the bottom (top).}. 
For $s_\beta=0$ the SM Higgs lives in the bulk and has KK-excitations,
whereas for $s_\beta=1$ it is a boundary field.
The $W$ gauge boson mass matrix is given by
\begin{equation}
{\cal M}^2_W\simeq \pmatrix{m^2_W&\sqrt{2}m^2_Ws^2_\beta
&\sqrt{2}m^2_Ws^2_\beta&\dots\cr
 \sqrt{2}m^2_Ws^2_\beta&M_c^2&&\cr
  \sqrt{2}m^2_Ws^2_\beta&&(2M_c)^2&\cr
 \vdots &&&\ddots\cr}\, ,
\label{mass}
\end{equation}
where $M_c\equiv 1/R$, $m^2_W=g^2v^2/2$, and $g$ is the SU(2)$_L$ 
gauge coupling. 
In Eq.~(\ref{mass}) we have  neglected terms of 
$\calo (m^2_W)$ for the KK-excitation masses, since they are subleading
in the limit $M_c\gg m_W$ considered here.
From now on we will only consider the leading
corrections, of $\calo (m^2_W/M_c^2)$, to the masses and couplings of
the lightest modes.
The eigenvalues of the matrix (\ref{mass}) can be obtained 
at this order by the rotation ${\cal  R} {\cal M}^{2}_W{\cal R}^\dagger$,
with
\begin{equation}
{\cal R}\simeq \pmatrix{1&\theta_1&\theta_2&\dots\cr
                         -\theta_1&1&&\cr
                         -\theta_2&&1&\cr
                        \vdots &&&\ddots\cr}\ ,\ \ 
 {\rm and}\ \ \ \ \
\theta_n=-\frac{\sqrt{2}m^2_Ws^2_\beta}{n^2M_c^2}\, .
\label{rotation}
\end{equation}
The mass eigenvalues are
\begin{eqnarray}
\mwph&=&m^2_W\left[1-2
s_\beta^4
\sum^\infty_{n=1}\frac{m^2_W}{n^2M_c^2}\right]
\, ,\label{massw}\\
M^{(n)\, 2}_{KK}&=&
n^2M_c^2+{\cal O}\left(m^2_W\right)\ ,\ \   n=1,2,...,\infty\, . 
\end{eqnarray}
The lightest mode, of mass $\mwph$,
is the one to be associated with the SM $W$ boson.
Its  coupling to the fermions is affected by the rotation (\ref{rotation}).
We obtain
\begin{equation}
\gph=
g\left[1-2
s_\beta^2\sum^\infty_{n=1}\frac{m^2_W}{n^2M_c^2}\right]
\, .
\label{coupling}
\end{equation}
For the neutral SM gauge bosons,  $W_3$ and $B$, the situation is
analogous.
After the usual rotation by the electroweak angle $\theta_W$, the states
are split into the massless $\gamma$ 
plus its KK-excitations  (with masses  $nM_c$),
and the KK-tower of $Z$s, 
whose mass matrix is identical to Eq.~(\ref{mass}) 
with the replacement $m_W\rightarrow m_Z$.
The lightest $Z$ boson has 
a mass and a gauge coupling to the fermions given by
\begin{eqnarray}
\mzph&=&m^2_Z\left[1-2
s_\beta^4\sum^\infty_{n=1}\frac{m^2_Z}
{n^2M_c^2}\right]\, ,\label{couplingz}\\
\gzph&=&
\frac{g}{\cos\theta_W}
\left[1-2
s_\beta^2\sum^\infty_{n=1}\frac{m^2_Z}{n^2M_c^2}\right]\, ,
\label{massz}
\end{eqnarray}
where $m_Z$ and $g/\cos\theta_W$ would be the mass and the coupling 
in the case of no mixing between the $Z$ and its KK-excitations.

\section{Electroweak observables and constraints on $M_c$}

Let us start considering
the effect of the KK-tower to the SM tree-level relation 
\begin{equation}
G^{SM}_F=
\frac{\pi\alpha}{{\sqrt{2}\mwph[1-\mwph/\mzph]}}\, ,
\label{GmuSM}
\end{equation}
where by $G_F^{SM}$ we refer to the SM prediction for
 the Fermi constant measured
in the $\mu$-decay, and 
$m_W^{(ph)}$ and $m_Z^{(ph)}$ are the measured (physical)
masses. 
In our model the $\mu$-decay can be also mediated by the 
$W$ excitations. Therefore we have
\begin{equation}
\frac{G_F}{\sqrt{2}}=\frac{\gphs}{8\mwph}+\sum^\infty_{n=1}
\frac{(\sqrt{2} g)^2}{8n^2M_c^2}\, ,
\label{GmuKK}
\end{equation}
where now $\gph$ and $\mwph$ are given in Eqs.~(\ref{coupling})
and (\ref{massw}), respectively.
On the other hand,
$\alpha$ is defined at zero momentum where
KK-contributions are negligible. Then we have
\begin{equation}
\alpha=\frac{g^2}{4\pi}\left(1-\frac{m^2_W}{m^2_Z}\right)
=\frac{\gphs}{4\pi}\left(1-\frac{\mwph}{\mzph}\right)
\left[1+(2s_\beta^4+4s_\beta^2)
\sum^\infty_{n=1}
\frac{\mwph}{n^2M_c^2}\right]\, .
\label{alpha}
\end{equation}
From Eqs.~(\ref{GmuKK}), (\ref{alpha}) and (\ref{GmuSM}),
we obtain
\begin{equation}
G_F=G^{SM}_F
\left[1-
2(s_\beta^4+2s_\beta^2-1)
\sum^\infty_{n=1}\frac{\mwph}{n^2M_c^2}
\right]\, ,
\label{Gmu}
\end{equation}
that expresses the deviation versus the SM prediction due to the 
KK-excitations.

In order to compare with the high-precision electroweak data,
we must include radiative corrections.
The loop effects of the KK-excitations can be neglected in the limit
$M_c\gg m_W$ 
\footnote{
These  loop effects, however, modify the gauge coupling
of the KK-excitations. We estimate that they can produce a $\lappeq 10\%$ 
of variation on 
the KK-contribution to the electroweak observables.}. 
In consequence we must only consider
the ordinary SM radiative corrections.
These can be easily 
incorporated by replacing the tree-level
relation (\ref{GmuSM}) by the loop-corrected one, that
can be extracted from Ref.~\cite{pdg}.
The excellent agreement between $G^{SM}_F$
and the observed value leads to a  severe 
constraint on the ratio $G_F/G^{SM}_F-1$.
Actually, since the experimental determination of $G_F$ is still
more precise than $\mwph$, the analysis of electroweak
observables uses $G_F$, $\mzph$ and $\alpha$ as input parameters,
and takes the relation in (\ref{GmuSM}),
corrected by radiative corrections (see Eq.~(10.6a) of Ref.~\cite{pdg}),
as a SM prediction for the
$W$  physical mass $m^{SM}_{W}$.
This must be compared with the
experimental value \cite{data} $m^{(ph)}_{W}=80.39\pm 0.06$ GeV. Using 
Eq.~(\ref{Gmu}) and the relation $\sum^{\infty}_{n=1} 1/n^2
=\pi^2/6$, we derive at the 2$\sigma$ level
\begin{equation}
{\mwphS [ \mzph-\mwphS ]\over {\mwph [ \mzph-\mwph ] }}=
\left[1+
(s_\beta^4+2s_\beta^2-1)\frac{\pi^2\mwph}{3M_c^2}
\right] = 1^{+0.0088}_{-0.0083}\, .
\label{mw}
\end{equation}
This translates into the lower bound on $M_c$
given in Fig.~1.
For $s_\beta=0$ we have $M_c\gappeq 1.6$ TeV. 
A similar bound was derived (in this limit $s_\beta=0$) 
in Ref.~\cite{nath}.
Notice that the bound depends strongly on $s_\beta$ and 
goes to zero for $s^2_\beta=\sqrt{2}-1$.
Therefore, we find that this is not a good observable to 
constrain $M_c$ in a model independent way.

We can proceed as above to obtain predictions for 
other electroweak observables.
We have considered three more quantities: (1) $Q_W$, 
obtained in atomic parity violating experiments \cite{pdg}, 
(2) $\Gamma (l^+l^-)$, the leptonic width of the $Z$, and
(3) the $\rho$-parameter defined as 
$\rho=m^{(ph)\, 2}_W/(m^{(ph)\, 2}_Z\cos^2\hat\theta_W)$
\cite{pdg}, where $\hat\theta_W$ 
is  the Weinberg angle 
in the $\overline{MS}$
scheme. 
The latter can be related \cite{pdg} to the physical quantity
$\sin^2\theta^{eff}_W
\equiv (1-g_V/g_A)/4$ that is measured at
 LEP and SLD \cite{data}.
KK-excitations do not contribute to the ratio $g_V/g_A$ and 
therefore do not modify $\sin^2\theta^{eff}_W$ nor 
$\sin^2\hat\theta_W$. 
The effect of the KK-excitations on  
$\Gamma(l^+l^-)$, $Q_W$ and $\rho$ is given by
\begin{eqnarray}
\Gamma (l^+l^-)
&=&\Gamma(l^+l^-)^{SM}
\left[1+
2\left[(s_\beta^2-1)^2\sin^2\theta_W-1\right]
\sum^\infty_{n=1}\frac{\mzph}{n^2M_c^2}
\right]
\label{gamma}\, ,\\
Q_W&=&Q^{SM}_W
\left[1+
2(s_\beta^2-1)^2\sin^2\theta_W
\sum^\infty_{n=1}\frac{\mzph}{n^2M_c^2}
\right]\, ,
\label{QW}\\
\rho&=&\rho^{SM}
\left[1-
2s^4_\beta\sin^2\theta_W
\sum^\infty_{n=1}\frac{\mzph}{n^2M_c^2}
\right]\, .
\label{rho}
\end{eqnarray}
where $\Gamma(l^+l^-)^{SM}$ is the SM prediction
written  as a function of  $G_F$, $m^{(ph)}_Z$
and $\sin^2\hat\theta_W$,  
whereas
$Q^{SM}_W$ is the SM prediction written as a function 
of $\sin^2\hat\theta_W$ 
and  $\rho^{SM}=1.0109\pm 0.0006$.
All these SM expressions can be found in Ref.~\cite{pdg}.
Comparing the predictions (\ref{gamma})-(\ref{rho})
with the experimental values, we  can place  bounds
on $M_c$ . 
These are shown in Fig.~1. 
The experimental values 
for $Q_W$ and $\Gamma(l^+l^-)$  have been taken
from Ref.~\cite{pdg}.
The experimental value of $\sin^2\hat\theta_W$ that appears in 
$\rho$ has been taken from Ref.~\cite{data}.
We find that 
the strongest lower bound on $M_c$ comes from the leptonic $Z$ width, an
observable that seems to be very appropriate
to constrain models with extra gauge bosons.
This is because (a) it is measured at the level of 0.1$\%$, (b)
the SM loop corrections are calculated with an even better precision
\footnote{Expressing $\Gamma (l^+l^-)$
as a function of $G_F$ and $m^{(ph)}_Z$ 
incorporates the largest radiative correction  from the running of
$\alpha$, and the dominant top correction appears
suppressed by a factor 
$\tan^2\theta_W$ with respect to that in Eq.~(\ref{GmuSM}).}
and (c)
its dependence on $s_\beta$ is very mild.
We find an absolute bound of $M_c\gappeq 2.5$ TeV.
The bound coming from $Q_W$ is much weaker. 
This disagrees with 
Ref.~\cite{nath}, where a stronger bound from $Q_W$
was obtained in the limit $s_\beta=0$.
We think that the reason of this
disagreement is that in
 Ref.~\cite{nath} $Q_W$ was derived not as a function of 
$Q_W^{SM}$  but as a function of $Q_W^{SM} G_F^{SM}/G_F$.
The bound from $\rho$ is not very strong either. This is  
due to the fact that
the gauge boson sector  has an approximate SU(2)-custodial symmetry 
only broken by the difference $(m^2_Z-m^2_W)/m^2_Z\simeq 0.23$.

One can look for other observables that would lead to analogous
bounds. For example, the total width of the $Z$ or 
$\sin^2\hat\theta_W$ from
the relation in Eq.~(10.9a) of Ref.~\cite{pdg}.
The latter also gives lower bounds around $2.5$ TeV, 
but with a strong dependence on $s_\beta$.

KK-excitations also affect the differential cross-sections for
 $e^+e^-\rightarrow f^+f^-$ measured at high energies, $q^2>m_Z^2$.
These experiments can be 
 used to test four-fermion contact interactions, and consequently 
to put lower bounds on the masses of the KK-excitations \cite{ab}.
We find that the largest bound comes from limits on 
the vector four-fermion interaction, 
$\epsilon_V[e^+\gamma^\mu e^-][f^+\gamma_\mu f^-]$.
In our model these are mediated (predominantly)
by the KK-tower of the photon and gives 
$\epsilon_V=-2q_f^2e^2\sum^\infty_{n=1}1/(n^2M^2_c)$.
The minus sign indicates that
the contribution interferes destructively with the SM one.
The strongest constraint on $\epsilon_V$ is found in the LEP2 experiment,
that gives  $\epsilon_V< 4\pi/{(9.3)^2}$ TeV$^{-2}$ 
for leptons at the 95$\%$ CL \cite{opal}.
This implies $M_c\gappeq 1.5$ TeV.
Constraints on $M_c$ can also be obtained from
direct searches for $Z^\prime$ 
at Tevatron \cite{tevatron}.
The present limit for a SM-like $Z^\prime$ is $M_{Z^\prime}>690$ GeV.
In our model, however, we must consider that the
coupling of the KK-excitations to fermions 
is a factor 
$\sqrt{2}$ larger than that of the $Z$, and 
the cross-section production is enhanced by a factor of 4. 
From Ref.~\cite{tevatron}, we get the  limit  $M_c\gappeq 820$ GeV.
Similarly, from searches for $W^\prime$ \cite{tevatronb} we obtain 
the bound $M_c\gappeq 780$ GeV.

Finally, we want to comment on 
models with more than one extra dimension.
Our analysis can be easily extended just by replacing
the sum $\sum^{\infty}_{n=1}1/n^2$ appearing in the above equations
by the sum over all the KK-excitations of the theory.
This sum, however,  
depends on the manifold on which the theory is
compactified \cite{ab}.
In addition, for more than one extra dimension it
is not finite. For two extra dimensions, for example,
the sum diverges logarithmically $\sim\ln(\Lambda/M_c)$ and 
therefore depends also on the cutoff of the theory $\Lambda$. 
Consequently, the lower bounds on $M_c$ 
for more than one extra dimension will be stronger 
but very model dependent.

\section{Conclusions} 

There are well motivated theoretical arguments
that imply the existence of more than three spatial dimensions.
In order to be consistent with all observations, of course,
the extra dimensions must be compactified 
at some high-energy scale. If this scale is around the TeV, 
their presence must affect the SM electroweak 
predictions currently being tested at 
high precision experiments.

In this letter we have analyzed these effects.
We have shown 
how the associated tower of KK-excitations of the SM 
fields modify the relations between different electroweak observables.
We have considered the most general case by taking the SM Higgs
doublet as a combination of a field
living in the 5D bulk 
and another  living on the 4D boundary of the manifold. 
We have compared 
with the present electroweak data and have put 
constraints on the compactification scale.
We have shown that, if an extra dimension exists, 
it must be compactified at a scale larger than $\approx 2.5$  TeV.
This bound will
 be improved with a better 
experimental determination of, for example, the $W$ mass. 
Also new LEP2 data on differential cross-sections for 
$e^+e^-\rightarrow f^+f^-$ can, as discussed above, be very 
useful to establish the maximum length of an extra dimension.

\section*{Acknowledgements} 

We thank Masahiro Yamaguchi for discussions.
The work of M.M. was supported by CICYT under contract AEN96-1672
and by the Junta de Andaluc\'\i a under contract FQM-101.

\newpage

\setlength{\unitlength}{1cm}
\begin{figure}[htb]
\begin{picture}(10,10)
\epsfxsize=20cm
\put(-2.,-17){\epsfbox{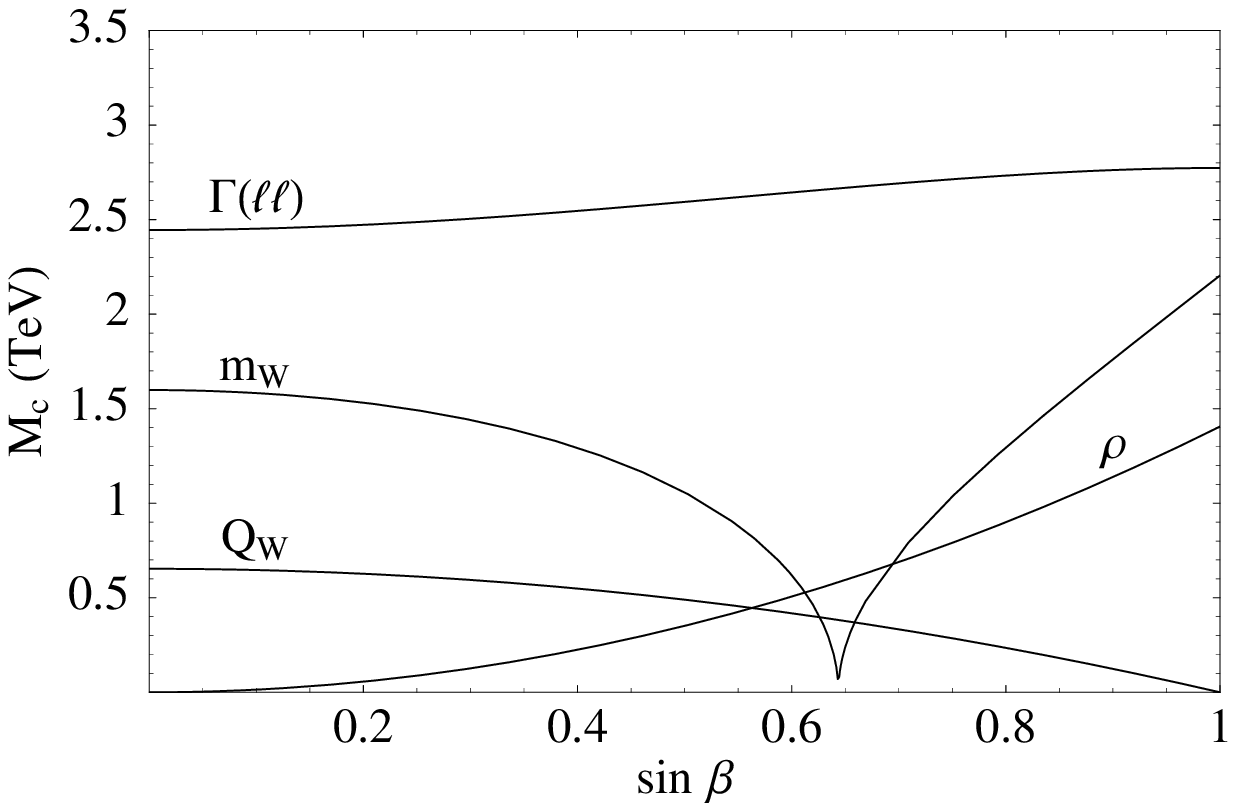}}
\end{picture}
\caption{Lower bounds on the compactification scale 
$M_c$ from electroweak observables.
\label{Fig. l}}
\end{figure}

\end{document}